\documentstyle[amssymb,aps,prb,twocolumn,psfig]{revtex}

\begin{document}

\twocolumn[\hsize\textwidth\columnwidth\hsize\csname@twocolumnfalse\endcsname 
]

{\bf Schopohl and Dolgov reply: }let us first summarize the pertinent result
of the subject letter\cite{Schopohl and Dolgov}, before addressing
statements made in the preceding interesting and stimulating comments\cite
{Volovik}, \cite{Hirschfeld et al.}. We\cite{Schopohl and Dolgov} consider
the {\em total} electromagnetic contribution to the free energy ${\cal F}$
for a {\em fixed} (externally controlled) current distribution ${\bf j}%
_{ext}({\bf q)}$ in a superconductor. To ease comparison of our conclusions%
\cite{Schopohl and Dolgov} with the foundational comment of Volovik\cite
{Volovik} we rewrite ${\cal F}$ in the equivalent form\cite{Schopohl and
Dolgov} 
\begin{equation}
{\cal F}=-\frac{2\pi }{c^{2}}\int \frac{d^{3}q}{(2\pi )^{3}}\frac{\left| 
{\bf j}_{ext}\left( {\bf q}\right) \right| ^{2}}{q^{2}+\frac{1}{\lambda ^{2}(%
{\bf q,}T)}}  \label{free energy}
\end{equation}
where $\lambda ({\bf q,}T)$ is the temperature dependent magnetic
penetration depth ({\bf an eigenvalue of the electromagnetic kernel}). From
this one readily calculates the electromagnetic part of the entropy $S(T)=-%
\frac{\partial {\cal F}}{\partial T}$ for any externally controlled ({\em %
finite}) current distribution ${\bf j}_{ext}{\bf (q)}$: 
\begin{equation}
S(T)=-\frac{2\pi }{c^{2}}\int \frac{d^{3}q}{(2\pi )^{3}}\frac{\partial }{%
\partial T}\left[ \frac{1}{\lambda ^{2}({\bf q,}T)}\right] \cdot \frac{%
\left| {\bf j}_{ext}\left( {\bf q}\right) \right| ^{2}}{\left[ q^{2}+\frac{1%
}{\lambda ^{2}({\bf q,}T)}\right] ^{2}}
\end{equation}
We argued\cite{Schopohl and Dolgov}, that a strictly linear $T$-dependence
of the magnetic penetration depth (MPD) in a superconductor, of the form $%
\lambda (T)-\lambda (0)\propto T$, violates indeed the third law of
thermodynamics.

Volovik\cite{Volovik} discusses a situation, where a stationary flow
(associated with an {\em uncharged} superfluid) circulates in a ring. He
also suggests\cite{Volovik} to compare a charged system with an uncharged
one: the externally controlled current density ${\bf j}_{ext}$ should be
related to the superfluid velocity ${\bf v}_{s}$, the magnetic penetration
depth $\lambda (T{\bf )}$ should be identified with the superfluid density $%
\rho _{s}(T)=\rho -\rho _{n}(T)$, the free energy ${\cal F}$ in Eq.(\ref
{free energy}) should be equivalent to the kinetic energy $\frac{1}{2}\rho
_{s}(T)\,{\bf v}_{s}^{2}$ of the flow of a neutral superfluid\cite{Volovik}.
However, such an identification is deceptively simple:

a) ${\bf j}_{ext}\ $is externally controlled, and henceforth strictly
independent on $T$, while ${\bf v}_{s}$, as discussed by Volovik, may in
principle display a $T$-dependence. In some cases the $T$-dependence of $%
{\bf v}_{s}$ may be irrelevant, in other cases it may be important,
depending on circumstances how ${\bf v}_{s}$ is measured. For example, what
is the meaning of ${\bf v}_{s}$ in a microwave absorption experiment?

b) ${\cal F}$ describes a lot more than just the kinetic energy of the
superflow, since ${\cal F}$ explicitely takes into account the full
electromagnetic interaction of the ordered medium (in our case the
superconductor) with the externally controlled $c$-number source ${\bf j}%
_{ext}$ (see Ref.\cite{Kirzhnitz} for an authoritative discussion of the
electrodynamics of ordered macroscopic media). Also, Volovik restricts his
discussion to the hydrodynamic limit $\left| {\bf q}\right| \rightarrow 0$,
while ${\cal F}$ takes into account also the non local effects, important
for wavevectors around $\left| {\bf q}\right| \cdot \lambda (T)\gtrsim 1$.

c) Volovik states\cite{Volovik} that the total current density ${\bf j}$
should be proportional to ${\bf v}_{s}$, via ${\bf j=}\rho _{s}{\bf v}_{s}$,
so that the 'response' $\rho _{s}$ is defined in the limit $\left| {\bf v}%
_{s}\right| \rightarrow 0$, while the Nernst principle requires first $%
T\rightarrow 0$ at finite ${\bf v}_{s}$. Should we distrust electromagnetic
linear response theory? Can it not be applied to a charged superfluid at low
temperatures when there exists a finite superflow in the ground state? Our
answer is that the transversal electromagnetic linear response function\cite
{Kirzhnitz}, defined via \cite{Schopohl and Dolgov} 
\begin{equation}
\left[ {\bf q}^{2}-\frac{\omega ^{2}}{c^{2}}\varepsilon _{tr}({\bf q,}\omega
)\right] {\bf A(q\,,}\omega )=\frac{4\pi }{c}{\bf j}_{ext}{\bf (q,}\omega )
\end{equation}
is physically correct for any externally controlled $c$-number current ${\bf %
j}_{ext}$, provided $\left| {\bf j}_{ext}\right| $ is {\em small} compared
to the microscopic atomic (molecular) currents in the matter. Indeed, the
physical notion and subsequent distinction between (weak) {\em external}
fields and (possibly strong) {\em local} fields inside the matter adequately
addresses this point. Only under extreme conditions external fields become,
eventually, comparable to the strength of local fields, for example in laser
physics. Within the range of validity of linear optics, however, the linear
response of a superconductor, represented by the {\em transversal}
dielectric function $\varepsilon _{tr}$, should be physically adequate to
describe the reaction of a superconductor to any externally controlled
electromagnetic influence ${\bf j}_{ext}$, and this at all temperatures $T$.

This view is also taken by Hirschfeld et al.\cite{Hirschfeld et al.}. These
authors discuss the static ${\bf q}$-dependent electromagnetic response
kernel $\lim_{\omega \rightarrow 0}\frac{\omega ^{2}}{c^{2}}\left[ 1-%
\mathop{\rm Re}%
\varepsilon _{tr}({\bf q,}\omega )\right] $ , which they construct (using
standard linear response theory) out of a product of two Matsubara Green's
functions having as its source the weak coupling BCS theory. Their
electromagnetic kernel applies for conventional as well as unconventional
pairing symmetry. The interesting point made by Hirschfeld et al. is, that
irrespective of the orientation of ${\bf q}$ with respect to the $ab$-plane
of a layered superconductor, there exists a tiny correction of order $q^{2}$%
, that should stabilize a pure $d_{x^{2}-y^{2}}$ symmetry BCS pairing
groundstate at low temperatures, even in the exceptional case\cite{Schopohl
and Dolgov} when ${\bf q\cdot v}_{F}=0$ . They estimate the crossover
temperature $T^{*}$, below which a deviation of the linear $T-$dependence of
MPD should be detected, around $k_{B}T^{*}\simeq \frac{\hbar ^{2}\omega
_{pl}^{2}}{2mc^{2}}$. Such a crossover temperature $T^{*}$ might be outside
the predictive power of the (non relativistic) electromagnetic response
theory, though. But accepting their argument, the electromagnetic kernel
becomes intrinsically non local. This leads trivially to a non linear $T$%
-dependence of MPD fulfilling $\lim_{T\rightarrow 0}\frac{\partial \lambda (%
{\bf q,}T{\bf )}}{\partial T}=0$, in accordance with the Nernst principle%
\cite{Schopohl and Dolgov}. We agree: a $d_{x^{2}-y^{2}}$-pairing
groundstate may be stable at low temperatures, but it will {\em not} display
a linear $T-$dependence of MPD due to non local effects \cite{leggett},\cite
{Hirschfeld et al.}. For $T\rightarrow 0$ the conceptual distinction between
a {\em non local} electromagnetic response kernel on one hand and a {\em %
local} superfluid density $\rho _{s}$ on the other hand becomes important.
The answer to the question posed in the title of the subject letter\cite
{Schopohl and Dolgov} is no.

N. Schopohl and O.V. Dolgov

Eberhard-Karls-Universit\"{a}t T\"{u}bingen

Institut f\"{u}r Theoretische Physik

Auf der Morgenstelle 14

D-72076 T\"{u}bingen, Germany

Received xxx June 1998


\vspace{5cm}


\begin{references}
\bibitem{Schopohl and Dolgov}  N. Schopohl and O.V. Dolgov, \prl  {\bf 80},
4761(1998).

\bibitem{Volovik}  G.E. Volovik, cond-mat/9805159.

\bibitem{Hirschfeld et al.}  P.J. Hirschfeld, M.-R. Li and P. W\"{o}lfle,
cond-mat/9806085.

\bibitem{leggett}  I. Kosztin and A.J. Leggett, \prl {\bf 79}, 135, (1997).

\bibitem{Kirzhnitz}  D.A. Kirzhnits ''General Properties of Electromagnetic
Response Functions'', Ch. 2, in: {\it The Dielectric Function of Condensed
Systems}, eds., L.V. Keldysh, D.A. Kirzhnits, A.A. Maradudin, Elsevier Publ.
(1989).
\end{references}
\end{document}